\documentclass[twocolumn,showpacs,pra,aps,superscriptaddress]{revtex4}
\usepackage{ulem}
\usepackage{array}
\usepackage{amsmath}
\usepackage{graphicx}
\usepackage{dsfont}
\usepackage{color}
\def\bra#1{\langle #1|}
\def\ket#1{|#1 \rangle}

\begin{document}

\title{Maximal violation of tight Bell inequalities
for maximal high-dimensional entanglement}

\author{Seung-Woo Lee}

\affiliation{Clarendon Laboratory, University of Oxford, Parks Road,
Oxford OX1 3PU, United Kingdom}

\author{Dieter Jaksch}

\affiliation{Clarendon Laboratory, University of Oxford, Parks Road,
Oxford OX1 3PU, United Kingdom}\affiliation{Center for Quantum
Technologies, National University of Singapore, Singapore 117543,
Singapore}

\affiliation{Keble College, Parks Road, Oxford OX1 3PG, United
Kingdom}

\date{\today}

\begin{abstract}
We propose a Bell inequality for high-dimensional bipartite systems
obtained by binning local measurement outcomes and show that it is
tight. We find a binning method for even $d$-dimensional measurement
outcomes for which this Bell inequality is maximally violated by
maximally entangled states. Furthermore we demonstrate that the Bell
inequality is applicable to continuous variable systems and yields
strong violations for two mode squeezed states.
\end{abstract}


\maketitle

%\section{Introduction}
% Introduction %

The incompatibility of quantum non-locality with local-realistic
(LR) theories is one of the most remarkable aspects of quantum
theory. LR theories impose constraints on the correlations between
measurement outcomes on two separated systems which are described by
Bell inequalities (BIs) \cite{Bell64}. It was shown that BIs are
violated by quantum mechanics in the case of entangled states.
Therefore BIs are of great importance for understanding the
conceptual foundations of quantum theory and also for investigating
quantum entanglement. Since the first discussion of quantum
non-locality by Einstein-Podolski-Rosen (EPR) a great amount of
relevant work has been done and numerous versions of BIs have been
proposed
\cite{Bell64,CHSH69,Collins02,Acin02,WSon05,SWLee07,Masanes03,
Reid88,Chen02,Jeong03,Banaszek99,Braunstein05,Gisin07}.

For bipartite $2$-dimensional systems the CHSH Bell-type inequality
\cite{CHSH69} has the desirable property of only being maximally
violated for a maximally entangled state. The CHSH inequality
divides the space of correlations between measurement outcomes by
defining a hyperplane. Since a facet of the polytope defining the
region of LR correlations lies in this hyperplane the CHSH
inequality is tight. This means that any violation of LR theories
occurring on this particular facet is indicated by the CHSH
inequality \cite{Masanes03}. Tightness is a desirable property since
only sets of tight BIs can provide necessary and sufficient
conditions for the detection of pure state entanglement. There are
still many open questions regarding the generalization of BIs to
complex quantum systems \cite{Gisin07}. For example, BIs for
bipartite high-dimensional systems as e.g.~that proposed by Collins
{\em et al.} \cite{Collins02} are either not maximally violated by
maximal entanglement \cite{Acin02} or as in the case of Son {\em et
al.} \cite{WSon05} were shown to be non-tight \cite{SWLee07}.

In the case of continuous variable systems there is so far no known
BI formulated in phase space which is maximally violated by the EPR
state -- the maximally entangled state associated with position and
momentum \cite{Braunstein05}. Although Banaszek and Wodkievicz (BW)
showed how to demonstrate non-locality in phase space
\cite{Banaszek99} their BI is not maximally violated by the EPR
state \cite{Jeong03}. Another approach using pseudospin operators
was shown to yield maximal violation for the EPR state
\cite{Chen02}. However, finding measurable local observables to
realise this approach is challenging. Due to the lack of any known
BI providing answers to these questions we still have no clear
understanding of nonlocal properties of high-dimensional systems and
their relation to quantum entanglement.

In this paper we present a BI for even $d$-dimensional bipartite
quantum systems which, in contrast to previously known BIs, fulfills
the two desirable properties of being tight and being maximally
violated by maximally entangled states. These properties are
essential to investigate quantum non-locality appropriately and for
consistency with the $2$-dimensional case. We call BIs fulfilling
these properties {\em optimal BIs} throughout this paper. Then we
extend optimal BIs to continuous variable systems and demonstrate
strong violations for properly chosen local measurements.

{\em Optimal Bell inequality--}We begin by briefly introducing the
generalized formalism for deriving BIs for arbitrary $d$-dimensional
bipartite systems \cite{SWLee07}. Suppose that two parties, Alice
and Bob, independently choose one of two observables $\hat{A}_1$ or
$\hat{A}_2$ for Alice, and $\hat{B}_1$ or $\hat{B}_2$ for Bob.
Possible measurement outcomes are denoted by $k_a$ for $\hat{A}_a$
and $l_a$ for $\hat{B}_b$ with $a,b=1,2$, where $k_a,l_b\in
V\equiv\{0,1,...,d-1\}$. A general Bell function is then written as
\cite{SWLee07}
\begin{eqnarray}
  \label{eq:GeneralBF}
  {\cal B}=\sum_{a,b=1}^{2}\sum_{k_a,l_b=0}^{d-1}\epsilon_{ab}(k_a,l_b)P_{ab}(k_a,l_b),
\end{eqnarray}
where $P_{ab}(k_a,l_b)$ is the joint probability for outcomes $k_a$
and $l_b$, and $\epsilon_{ab}(k_a,l_b)$ are their weighting
coefficients (here assumed to be real). For local-realistic (LR)
systems each probabilistic expectation of ${\cal B}$ is a convex
combination of all possible deterministic values. It can thus not
exceed the maximal deterministic expectation value given by
\begin{eqnarray}
  \label{eq:LRmax}
{\cal
B}^{\mathrm{max}}_{\mathrm{LR}}=\max_{C}\biggl\{\sum_{a,b=1}^{2}\epsilon_{ab}(k_a,l_b)\biggr\},
\end{eqnarray}
where $C\equiv\{(k_1,k_2,l_1,l_2)| k_1,k_2,l_1,l_2 \in V \}$ is the
set of all possible outcome configurations. A quantum state violates
local realism if its expectation value exceeds the bound ${\cal
B}^{\mathrm{max}}_{\mathrm{LR}}$. The flexibility in choosing the
coefficients $\epsilon_{ab}(k_a,l_b)$ allows the derivation of all
previously known BIs \cite{SWLee07} from Eq.~(\ref{eq:GeneralBF}),
e.g.~those proposed by Collins {\em et al.} \cite{Collins02} and by
Son {\em et al.} \cite{WSon05}. Moreover, we can construct new BIs
by properly choosing coefficients $\epsilon_{ab}(k_a,l_b)$. Our aim
is to find optimal BIs which fulfil the following conditions:

(C1) The BI is tight i.e.\ it defines a facet of the polytope
separating LR from non-local quantum regions in correlation or joint
probability space.

(C2) The BI is maximally violated by a maximally entangled state.
For each bipartite $d$-dimensional maximally entangled state there
exists a basis $\ket{j}$ with $j=0,\cdots,d-1$ in which this state
reads
$\ket{\psi_d^{\mathrm{max}}}=\sum^{d-1}_{j=0}\ket{jj}/\sqrt{d}$.

As a general method, one could choose the coefficients
$\epsilon_{ab}(k_a,l_b)$ freely and examine whether the resulting BI
satisfies the conditions (C1) and (C2). Here we instead propose a
method which restricts this choice and is guaranteed to give tight
BIs. We assume that the coefficients are products of arbitrary
binning functions defined by each party as
\begin{equation}
\zeta_{R}(k)=
\begin{cases}
+1& \text{if $\mathrm{outcome}~k\in R$},\\
-1& \text{otherwise},
\end{cases}
\end{equation}
where $R$ is an arbitrarily chosen subset of all possible outcomes,
i.e.\ $R\subset V$. The coefficients are then given by
\begin{eqnarray}
  \label{eq:Regcoeff2}
  \nonumber
\epsilon_{11}=\zeta_{R_1}(k_1)\zeta_{S_1}(l_1),~
\epsilon_{12}=\zeta_{R_1}(k_1)\zeta_{S_2}(l_2),\\
\epsilon_{21}=\zeta_{R_2}(k_2)\zeta_{S_1}(l_1),~
\epsilon_{22}=-\zeta_{R_2}(k_2)\zeta_{S_2}(l_2),
\end{eqnarray}
where $R_a$ and $S_b$ are subsets of the outcomes of $\hat{A}_a$ and
$\hat{B}_b$, respectively. From Eq.~(\ref{eq:LRmax}) we find the LR
upper bound ${\cal B}^{\mathrm{max}}_{\mathrm{LR}}=2$.

We first show that any BI derived by this method is tight. The
extremal points of the polytope separating LR and non-local quantum
mechanical correlations are associated with all deterministic
configurations $C$. They are described by $4d^2$ dimensional
linearly independent vectors
$\mathbf{G}_{k_1,k_2,l_1,l_2}=(\mathbf{e}_{k_1}\otimes
\mathbf{e}_{l_1})\oplus(\mathbf{e}_{k_1}\otimes
\mathbf{e}_{l_2})\oplus(\mathbf{e}_{k_2}\otimes
\mathbf{e}_{l_1})\oplus(\mathbf{e}_{k_2}\otimes \mathbf{e}_{l_2})$,
where $\mathbf{e}_{k}$ is the $d$-dimensional vector whose $k$-th
component is $1$ and all other components are zero. The interior
points of the polytope are given by convex combinations of these
extremal points and represent the region accessible to LR theories.
We now only consider extremal points associated with configurations
giving the maximal LR value ${\cal B}^{\mathrm{max}}_{\mathrm{LR}}$
and denote their number by $M$. For a polytope defined in $4d^2$
dimensions at least $4d(d-1)$ linearly independent vectors are
required to define a facet. Therefore, if $M \geq 4d(d-1)$ the
extremal points yielding ${\cal B}^{\mathrm{max}}_{\mathrm{LR}}$
define a facet of the polytope distinguishing LR from non-local
quantum mechanical correlations \cite{Masanes03}. We assume the
number of elements in the sets $R_1$, $R_2$, $S_1$, $S_2$ to be
$n_1$, $n_2$, $m_1$, $m_2$, respectively, where $0 \leq n_1, n_2,
m_1, m_2 \leq d-1$. We then count the number of configuration giving
${\cal B}^{\mathrm{max}}_{\mathrm{LR}}$ and find
$M=d^2(d^2-d(n_1+m_1)+n_1(m_1+m_2)+n_2(m_1-m_2))\geq 4d(d-1)$.
Therefore all BIs obtained by this method are tight, i.e.\ they
satisfy condition (C1). Note that any loss of elements in binned
subsets may cause them to become non-tight.

\begin{figure}
\begin{center}
\includegraphics[width=0.45\textwidth]{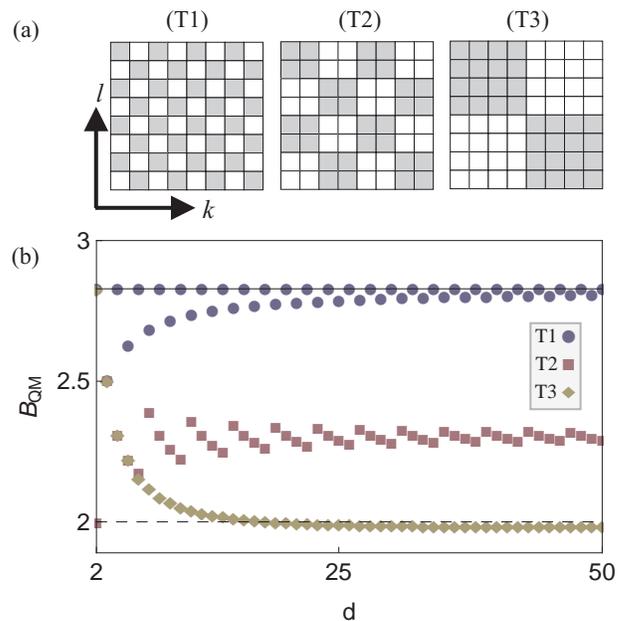}
\caption{(a) The coefficient distributions of BIs (T1), (T2), and
(T3) for $d=8$ are shown with weighting +1 (white for
$\epsilon_{11}$,$\epsilon_{12}$,$\epsilon_{21}$ and grey for
$\epsilon_{22}$) and -1 (grey for
$\epsilon_{11}$,$\epsilon_{12}$,$\epsilon_{21}$ and white for
$\epsilon_{22}$) in outcome space. (b) Quantum expectation values
${\cal B}_{\mathrm{QM}}$ of (T1), (T2), and (T3) are plotted. As $d$
increases, the expectation value of (T1) reaches the bound
$2\sqrt{2}$ (solid line), while that of (T2) approaches
$2.31<2\sqrt{2}$ and that of (T3) decreases below the
local-realistic upper bound 2 (dashed line).}\label{fig1}
\end{center}
\end{figure}

We now discuss the maximal violation condition (C2) by considering
three different tight BIs obtained via the above method. The
corresponding choices of the coefficients $\epsilon_{ab}$ for $d=8$
are schematically shown in Fig.~\ref{fig1}(a): (T1) is the sharp
binning type which can be realised if all outcomes are identifiable
with perfect measurement resolution. The elements of the subsets are
given as the even-numbers, i.e.\ $R_1=R_2=S_1=S_2=\{0,2,4,...\}$ so
that the coefficients $\epsilon_{ab}$ have an alternating weight
$+1$ or $-1$ when an outcome changes by one. (T2) is associated with
unsharp binning resolution and can be used to model imperfect
measurement resolution. The subsets are chosen as
$R_1=R_2=S_1=S_2=\{\forall k|k\equiv 0,1 (\mathrm{mod}~4)\}$ where
$k\equiv0,1(\mathrm{mod}~4)$ indicates that $k$ is congruent to $0$
or $1$ modulo $4$. The coefficients $\epsilon_{ab}$ alternate
between $+1$ and $-1$ for every $2$ outcomes. In type (T3) the
measurement results are classified into two divided regions by the
mean outcome $[d/2]$, where $[x]$ denotes the integer part of $x$.
The subsets are chosen as $R_1=R_2=S_1=S_2=\{\forall k|0\leq k<
[d/2]\}$. These three types of binning correspond to different
capabilities in carrying out measurements on $d$-dimensional
systems. Their properties will yield useful insights for testing BIs
in high dimensional and continuous variable systems.

We examine quantum violations of (T1), (T2) and (T3) by the
maximally entangled state $\ket{\psi_d^{\mathrm{max}}}$ with
increasing dimension $d$. The measurements $\hat{A}_a$ and
$\hat{B}_b$ are performed in the bases $\ket{a,
k}=(1/\sqrt{d})\sum_{j=0}^{d-1}\omega^{(k+\alpha_a)j}\ket{j}$ and
$\ket{b,
l}=(1/\sqrt{d})\sum_{j=0}^{d-1}\omega^{(l+\beta_b)j}\ket{j}$,
obtained by quantum Fourier transformation and phase shift
operations on $\ket{j}$. Here $\omega=\exp(2\pi i/d)$, and
$\alpha_a$ and $\beta_b$ are phase factors differentiating the
observables of each party $\hat{A}_a$ and $\hat{B}_b$, respectively.
The expectation value of the Bell function is then given by
\begin{eqnarray}
{\cal B}_{\mathrm{QM}}=\sum_{a,b=1}^{2}\sum_{k,l=0}^{d-1}
\frac{\epsilon_{ab}(k,l)}{2d^3\sin{[\frac{\pi}{d}(k+l+\alpha_a+\beta_b)]}}.
\end{eqnarray}
As shown in Fig.~\ref{fig1}(b), the expectation values of (T1) for
even-dimensions are $2\sqrt{2}$, and those for odd-dimensions tend
towards $2\sqrt{2}$ with increasing $d$. This is the upper bound for
quantum mechanical correlations which we show by defining a Bell
operator as ${\cal
\hat{B}}=\sum_{a,b}\sum_{k,l}\epsilon_{ab}(k,l)\ket{a,k}\bra{a,k}\otimes
\ket{b,l}\bra{b,l}$. From Eq.~(\ref{eq:Regcoeff2}), ${\cal
\hat{B}}^2=4\openone_d\otimes\openone_d+[\hat{P}_1,
\hat{P}_2]\otimes[\hat{Q}_2, \hat{Q}_1]$ where $\hat{P}_a=\sum_{k}
\zeta_{R_a}(k)\ket{a,k}\bra{a,k}$, $\hat{Q}_b=\sum_{l}
\zeta_{S_b}(l)\ket{b,l}\bra{b,l}$ and $\openone_d$ is the
$d$-dimensional identity operator. Since
$\|[\hat{P}_1,\hat{P}_2]\|\leq
\|\hat{P}_1\hat{P}_2\|+\|\hat{P}_2\hat{P}_1\|\leq2\|\hat{P}_1\|
\|\hat{P}_2\|=2$ and likewise for $\|[\hat{Q}_2,\hat{Q}_1]\|$ where
$\|\cdot\|$ indicates the supremum norm, we finally obtain $\|{\cal
\hat{B}}^2\|\leq 8$, or $\|{\cal \hat{B}}\|\leq 2\sqrt{2}$.

We calculate the quantum mechanical expectation value of ${\cal
\hat{B}}$ for BI (T1) by writing the coefficients as
$\epsilon_{11}=\epsilon_{12}=\epsilon_{21}=(-1)^{k+l}$ and
$\epsilon_{22}=-(-1)^{k+l}$. For even $d$ we use $\sum_{k,l=0}^{d-1}
(-1)^{k+l}/2d^3\sin{[\frac{\pi}{d}(k+l+\alpha_a+\beta_b)]}=
\cos{\pi(\alpha_a+\beta_b)}$ and find the expectation value
\begin{eqnarray}
 \label{eq:bellqm}
 \nonumber
{\cal
B}_{\mathrm{QM}}&=&\cos{\pi(\alpha_1+\beta_1)}+\cos{\pi(\alpha_1+\beta_2)}\\
&&\qquad+\cos{\pi(\alpha_2+\beta_1)}-\cos{\pi(\alpha_2+\beta_2)}.
\end{eqnarray}
This expression also holds approximately for sufficiently large odd
$d$. Thus we obtain ${\cal
  B}_{\mathrm{QM}}=2\sqrt{2}$, i.e.\ the maximal quantum upper
bound, for $\alpha_1=0$, $\alpha_2=1/2$, $\beta_1=-1/4$, and
$\beta_2=1/4$. Figure \ref{fig1}(b) also shows the maximal
expectation values of (T2) which are smaller than $2\sqrt{2}$ and
approach $\approx 2.31$ with increasing $d$. The maximal expectation
values of (T3) decrease below the local-realistic upper bound $2$
with increasing $d$.

These results show that (T1) is an optimal BI for even $d$ which
satisfies conditions (C1) and (C2). The optimal correlation operator
is then written as $\hat{E}_{ab}= \hat{\Pi}_a \otimes \hat{\Pi}_b$.
with the local measurement $\hat{\Pi}_a=
\sum_{k=0}^{d-1}(-1)^{k}\ket{a,k}\bra{a,k}$. Finally, we obtain the
optimal BI
\begin{eqnarray}
 \label{eq:optimalBI}
  {\cal B}=E_{11}+E_{12}+E_{21}-E_{22} \leq
  2,
\end{eqnarray}
where $E_{ab}=\langle \hat{E}_{ab} \rangle =
\sum_{k,l}(-1)^{k+l}P_{ab}(k,l)$ is the correlation function. Note
that for $d=2$ Eq.~(\ref{eq:optimalBI}) is equivalent to the CHSH
inequality \cite{CHSH69}. We have thus shown that the perfect sharp
binning of arbitrary even dimensional outcomes (T1) provides an
optimal BI, while the other binning methods (T2) and (T3) tend to
neglect quantum properties and do not show maximal violation for
maximally entangled states.

%\section{Continuous variable systems}

{\em Continuous variable systems--}We extend the optimal BI (T1) to
a continuous variable system and calculate its violation by a
two-mode squeezed state (TMSS). This state can, for instance, be
realised by non-degenerate optical parametric amplifiers
\cite{Reid88} in photonic systems. It is written as
$\ket{\mathrm{TMSS}}=\mathrm{sech}{r}\sum_{n=0}^{\infty}\tanh^{n}{r}\ket{n,n}$
where $r>0$ is the squeezing parameter and $\ket{n}$ are the number
states of each mode. In the infinite squeezing limit $r\rightarrow
\infty$, this becomes the normalized EPR state \cite{Banaszek99}.

When directly following the procedure of the finite dimensional case
two problems arise: First, we obtain the local measurement basis by
applying the quantum Fourier transformation to $\ket{n}$. This is
equivalent to the phase states
$\ket{\theta}=(1/\sqrt{2\pi})\sum_{n=0}^{\infty}\exp{(in\theta)}\ket{n}$
which are not orthogonal and not eigenstates of any hermitian
observable. Therefore no precise phase measurement can be carried
out. Second, a naive extension of the sharp binning method to the
continuous case is impossible. Note that any coarse-grained
measurement tends to lose quantum properties \cite{Kofler07} and
lead to non-tight BI tests. From the above results for unsharp and
regional binning we also do not expect strong violations by these
methods for the continuous variable system.

Let us consider the Pegg-Barnett phase state formalism \cite{Phase}.
We approximate the quantum phase by an orthonormal set of phase
states in a $s+1$-dimensional truncated space $\ket{\theta,
k}=(1/\sqrt{s+1})\sum_{n=0}^{s}\exp{(in\theta_k)}\ket{n}$ where
$\theta_k=\theta+2\pi k/(s+1)$ and $k=0,1,...,s$. Note that $s$ is a
cutoff parameter (assumed here to be an odd number) and in the limit
$s \rightarrow \infty$ there exists a $\theta_k$ arbitrarily close
to any given continuous phase. The correlation operator can then be
written as
$\hat{E}(\theta,\phi)=\hat{\Pi}(\theta)\otimes\hat{\Pi}(\phi)$ using
the {\em phase parity operator}
$\hat{\Pi}(\theta)=\sum_{k=0}^{s}(-1)^{k}\ket{\theta, k}\bra{\theta,
k}$.

We consider a truncated TMSS
$\ket{\psi_s}=(\mathrm{sech}{r}/\sqrt{1-\tanh^{2s+2}{r}})\sum_{n=0}^{s}\tanh^n{r}
\ket{n,n}$ which tends to the $s+1$-dimensional maximally entangled
state for $r \rightarrow \infty$ and to the TMSS for an infinite
cutoff, $s \rightarrow \infty$. The preparation of this state can
for instance be achieved by the optical state truncation method
\cite{Truncation}.

\begin{figure}
\begin{center}
\includegraphics[width=0.5\textwidth]{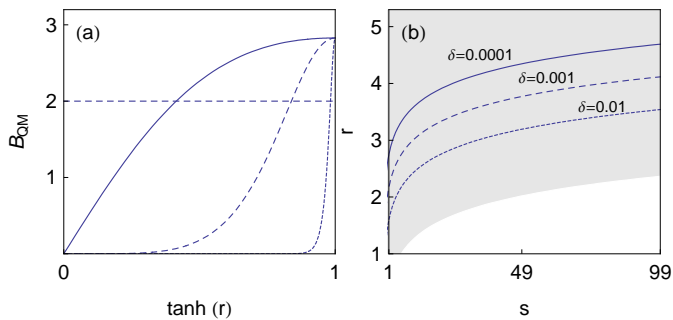}
\caption{(a) Expectation values of the Bell operator for truncated
TMSS with cutoff parameters $s=1$ (solid), $s=9$ (dashed), and
$s=99$ (dotted). (b) The shaded region indicates the values of $r$
for which the BI is violated; a violation better than ${\cal
B}_{\mathrm{QM}} \geq 2 \sqrt 2 - \delta$ occurs above the curves
shown for $\delta=0.01$ (dotted), $\delta=0.001$ (dashed), and
$\delta=0.0001$ (solid).}\label{fig2}
\end{center}
\end{figure}

The expectation value of the Bell operator is given by
\begin{eqnarray}
 \label{eq:Bexp}
{\cal B}_{\mathrm{QM}}&=&\bra{\psi_s}\hat{E}(\theta,
\phi)+\hat{E}(\theta, \phi')+\hat{E}(\theta', \phi)-\hat{E}(\theta', \phi')\ket{\psi_s} \nonumber \\
&=&4\sqrt{2}\frac{\tanh^{\frac{s+1}{2}}{r}}{1+\tanh^{s+1}{r}},
\end{eqnarray}
when $\theta=0$, $\theta'=\pi/(s+1)$, $\phi=-\pi/(2s+2)$, and
$\phi'=\pi/(2s+2)$. Fig.~\ref{fig2}(a) shows its monotonic increase
against the squeezing rate $r$ for different cutoff parameters $s$.
For any finite $s$ and $\delta > 0$ there exists a squeezing
parameter $r$ above which ${\cal B}_{\mathrm{QM}} \geq
2\sqrt{2}-\delta$. The required squeezing for this violation is $r
\geq \frac{1}{2}\ln[{(1+f(s,\delta))/(1-f(s,\delta))}]$ where
$f(s,\delta)=[(2\sqrt{2}-\sqrt{4\sqrt{2}\delta-\delta^2})/
(2\sqrt{2}-\delta)]^{2/(s+1)}$. The shaded region in
Fig.~\ref{fig2}(b) indicates the values of $r$ for which the BI is
violated ${\cal B}_{\mathrm{QM}}\geq 2$ and a violation better than
${\cal B}_{\mathrm{QM}} \geq 2 \sqrt 2 - \delta$ occurs for values
of $r$ above the corresponding curves for different $\delta$.
Violations arbitrarily close to the maximum value $2 \sqrt 2$ can
thus be achieved by sufficiently strongly squeezed states for any
finite value of $s$ with $r \rightarrow \infty$ corresponding to the
EPR state. Remarkably, this is in contrast to previous types of BIs
which were not able to get arbitrarily close to this bound for the
EPR state. However, we should note that for large $s$ one here again
faces difficulties in performing precise measurements due to the
indistinguishability of two local measurements as $\pi/(s+1)
\rightarrow 0$ for large $s$.

Finally, we discuss the relation of our optimal BI with the BW
inequality proposed in \cite{Banaszek99}. There, local measurements
are performed in the basis obtained by applying a Glauber
displacement operator $\hat{D}(\alpha)$ on the number states
$\ket{n}$. The measurement basis is written as
$\ket{\alpha,n}=\hat{D}(\alpha)\ket{n}$ with $\alpha$ an arbitrary
complex number. The displaced number operator is defined as
$\hat{n}_{\alpha}\equiv
\hat{D}(\alpha)\hat{n}\hat{D}^{\dag}(\alpha)$. Since
$\hat{n}_{\alpha}\ket{\alpha,n}= n\ket{\alpha,n}$, the correlation
operator is given by
$\hat{E}(\alpha,\beta)=\hat{\Pi}(\alpha)\otimes\hat{\Pi}(\beta)$,
where
$\hat{\Pi}(\alpha)=\sum_{n=0}^{\infty}(-1)^{n}\ket{\alpha,n}\bra{\alpha,n}$
is the displaced parity operator. Using this notation the BW
inequality becomes equivalent to Eq.~(\ref{eq:optimalBI}), which
shows that it is a tight BI for continuous variable systems.
However, the maximal expectation value of the BW inequality was
shown to be $2.32<2\sqrt{2}$ \cite{Jeong03}, while our type of BI
asymptotically reaches the bound $2\sqrt{2}$. This shows that the
optimal measurement bases for this non-locality test are obtained by
a quantum Fourier transformation on the standard bases \cite{Son04},
i.e.\ each of them is mutually unbiased to the standard basis. This
may also provide a useful insight about the optimality of measuring
in mutually unbiased bases for cases with more than two local
measurements~\cite{Ji08}.

% Conclusions %
%\section{Conclusions}

{\em Conclusions--}We derived, for the first time, a BI in even
$d$-dimensional bipartite systems which is maximally violated by
maximal entanglement and is also tight. These are desirable
properties for BIs in high-dimensional systems
\cite{Gisin07,SWLee07}. Our BI is found by perfectly sharp binning
of the local measurement outcomes. It can be used for testing
quantum non-locality for high dimensional systems, for instance it
coincides with the result for heteronuclear molecules by Milman {\em
et al.} \cite{Milman07}. Furthermore, we extended our studies to
continuous variable systems and demonstrated strong violations
asymptotically reaching the maximal bound $2\sqrt{2}$ for truncated
TMSSs by parity measurements in the Pegg-Barnett phase basis. This
provides a theoretical answer to the question of how maximal
violations of BIs can be demonstrated for the EPR states in phase
space formalism \cite{Braunstein05}. In the future we will
investigate the susceptibility of violations of our BIs to
measurement imperfections. In this context it will also be valuable
to search for additional optimal BIs comparing their properties and
extending optimal BIs to multipartite systems.

%\acknowledgments
We thank J. Lee, I. A. Walmsley and U. Dorner for
valuable discussions. This work was supported by the EU through the
STREP project OLAQUI and by the UK EPSRC through projects QIPIRC
(GR/S82176/01) and EuroQUAM (EP/E041612/1).

\end{document}